\documentclass[12pt]{article}
\usepackage{latexsym}
\usepackage{amssymb}
\usepackage{amsmath}
\usepackage{graphicx}

\begin{document}
\title{Wormhole throats in $R^{m}$ gravity}
\author {{\small N. Furey \footnote{e-mail: nmfurey@sfu.ca}} \\
\it{\small Department of Physics} \\ \it{\small Simon Fraser
University, Burnaby, British Columbia, Canada V5A 1S6}
 \and
{\small A. DeBenedictis \footnote{e-mail: adebened@sfu.ca}} \\
\it{\small Department of Physics} \\ \it{\small Simon Fraser
University, Burnaby, British Columbia, Canada V5A 1S6}}

\date{{\small November 12, 2004}}
\maketitle

\begin{abstract}
\noindent We consider wormhole geometries subject to a
gravitational action consisting of non-linear powers of the Ricci
scalar. Specifically, wormhole throats are studied in the case
where Einstein gravity is supplemented with a Ricci-squared and
inverse Ricci term. In this modified theory it is found that
static wormhole throats respecting the weak energy condition can
exist. The analysis is done locally in the vicinity of the
throat, which eliminates certain restrictions on the models introduced by considering the global topology.
\end{abstract}

\vspace{3mm}

\section{Introduction}
Wormhole type solutions in Einstein gravity have long been
studied. Flamm \cite{ref:flamm} seems to have been the first to
study such objects while considering the then newly discovered
solution of Schwarzschild. In 1935, Einstein and
Rosen considered wormhole type bridges as potential models for
elementary particles \cite{ref:einstrose}: the famous
Einstein-Rosen bridge. The field lay dormant for approximately two
decades until the consideration by Wheeler of the possibility of
a space-time foam. This foam, due to violent fluctuations in the
metric when considering quantum gravity, could be viewed as
wormhole-like structures permeating through space-time which would
potentially be manifest at energies near the Planck scale
\cite{ref:wheelfoam}. As well, wormholes have relevance to issues
such as chronology protection \cite{ref:chropro}, topology change
\cite{ref:topch} and in studies of horizons and singularities
\cite{ref:horiz}, \cite{ref:sing}.

\qquad There has been a sizeable amount of literature produced
over the past 15 years on the issue of Lorentzian as well as
Euclidean wormholes. This re-newed interest in the subject was
mainly sparked by the work of Morris and Thorne
\cite{ref:morthorn} where they considered static, traversable
wormholes for pedagogical purposes. In their work, and subsequent
papers by others, it was found that a static wormhole throat
could not be supported unless the weak energy condition (WEC) was
violated. The violation could be confined to a small region in
the vicinity of the throat but it must occur to prevent the
throat from collapsing (see \cite{ref:WEC1}, \cite{ref:WECmann},
\cite{ref:WECkuh}, \cite{ref:WECdeb}, \cite{ref:kar} and
references therein). Non-spherical geometries have also been
studied (some examples are \cite{ref:khat}, \cite{ref:eteo},
\cite{ref:rotberg}, \cite{ref:kuhax}).  For a more complete list
of pre-1996 references, as well as an excellent exposition on the
subject, the reader is referred to the book by Visser
\cite{ref:visbook}.

\qquad There have been many studies of static wormhole solutions
employing various exotic matter models within Einstein gravity.
However, it is possible that the action governing
gravitational dynamics is not the Einstein-Hilbert one, but
consists of a more complicated Lagrangian, which reduces to the
usual Einstein-Hilbert action in some limit. In this vein
theories with generalized actions of the form \footnote{Notations
and conventions here follow those of \cite{ref:MTW}}:
\begin{equation}
S=\frac{1}{16\pi}\int\,\mathcal{L}(R)\,\sqrt{g}\,d^{4}x +S_{matter}\; , \label{eq:genaction}
\end{equation}
with $\mathcal{L}(R)$ being some function of the Ricci scalar,
have been employed to explain various phenomena \cite{ref:nodin1}. The most popular
of these supplements the Einstein-Hilbert Lagrangian with an
$R^{2}$ term so that $\mathcal{L}(R)=R+\alpha\,R^{2}$ (for recent
work the reader is referred to \cite{ref:allem}) . This leads to
modifications which could be utilized to drive inflation purely
from the gravitational sector (the Starobinsky inflationary theory
\cite{ref:starob}). A thorough study of solutions in $R^{m}$ cosmology may be found in the recent papers \cite{ref:nodint2} as well as \cite{ref:cosdyn}.

\qquad A more recent modification which has received attention is
the addition of a $1/R$ term in the Lagrangian. The major
motivation for this modification is in the realm of matter
dominated era cosmology. For example, it is thought that perhaps
such a term may contribute to the present day acceleration of the
universe without the need for an exotic dark energy
\cite{ref:carroll}, \cite{ref:cct}, \cite{ref:mengwang}, \cite{ref:vollick}, \cite{ref:mengwang2}.
Inverse Ricci terms may also appear in certain sectors of string/M
theory (see \cite{ref:odin}, \cite{ref:gzbr} for example). These
are interesting ideas although it has been shown that adding
inverse powers of the Ricci scalar to the gravitational
Lagrangian is accompanied by some potential problems
\cite{ref:dolgov}, \cite{ref:chiba}, \cite{ref:doming}, \cite{ref:flan}.

\qquad Several issues are those of stability and the Newtonian
limit. It has been argued that, in both the metric variation and
the Palatini formulation of the $1/R$ theory, instabilities or
potentially unphysical weak-field limits may exist. The Newtonian
limit for the metric variation theory has been carefully studied
by Dick \cite{ref:rdick} who derived criteria for well defined
Newtonian limit. (A similar analysis has been done by
Dom\'{i}nguez and Barraco \cite{ref:doming} in the Palatini
formulation). We address this, and the issue of stability, later
in the paper. It has also been shown by Chiba \cite{ref:chiba}
that if the $1/R$ sector is to drive the present day cosmological
acceleration, the theory will yield results incompatible with
solar system experiments. However, this does not preclude
singular non-linear gravity theories which do not dominate
present day cosmological evolution yet whose effects may be
important under conditions as those potentially found near
wormhole throats. This is also the motivation in a recent paper
\cite{ref:CDDETT} where the future evolution of the universe is
considered with inverse curvature Lagrangians.

\qquad For the above reasons we believe it is of interest to study
wormhole geometries in a theory whose action is of the form
(\ref{eq:genaction}). That is, can the modified gravitational
sector of the theory eliminate the need for exotic matter in
supporting a static wormhole throat? The most complete case for
which the analysis is tractable seems to be the choice
$\mathcal{L}(R)=c_{-1}R^{-1} + R +c_{2}R^{2}$, the $c_{n}$'s
being the ``coupling'' constants of the theory. The $R^{2}$
theory was considered by Ghoroku and Soma \cite{ref:GS}. In their
meticulous study it was concluded that, under the assumption that
an asymptotically flat global solution exists, no WEC respecting
wormhole can exist in such a theory. The $1/R$ theory has not
been considered in the context of wormholes.

\qquad Here we study wormholes from a local geometric
perspective, without constraining the asymptotics. As pointed out
in \cite{ref:hochgeo}, geometric information is less limiting
than topology to the issue of defining and locating wormhole
throats. We find that in the local analysis, $R^{2}$ terms permit
the existence of a WEC respecting throat. The near throat
solution can be patched, via appropriate junction conditions, to
other solutions with desired far throat geometry and topology.
This approach is often taken in, for example, studies of stellar
structure or gravitational collapse. We briefly discuss the
junction conditions in a later section.

\qquad There are compelling reasons to believe that if
gravitational fields are dictated by an action such as
(\ref{eq:genaction}), then the non-linear contributions could be
important in the case of wormholes. The $R^{2}$ terms would
certainly be important in the high curvature regime which may be
present in the vicinity of wormhole throats. As well, higher
derivative contributions, introduced by both the $R^{2}$ and
$1/R$ terms, may be important since the wormhole structure only
places restrictions on the spatial components of the metric and
its first derivatives. We show below that, in the WEC, the
dominant terms near the throat are in fact those introduced by
these non-linear terms, regardless of how small the coupling
constants may be. It is found that the near throat region can
respect the WEC in the higher derivative theory.

\section{$\mathbf{R^{m}}$ gravity and the wormhole geometry}
The general $R^{m}$ action is given by:
\begin{equation}
S=\frac{1}{16\pi}\int d^{4}x\, \sqrt{g}\, \sum_{m} c_{m}R^{m}+ S_{matter}, \label{eq:polyaction}
\end{equation}
which, in the metric variation theory, gives rise to the equation of motion:
\begin{subequations}
\begin{align}
&R^{\mu}_{\;\nu}\sum_{m}mc_{m}R^{m-1} -\frac{1}{2}\delta^{\mu}_{\;\nu} \sum_{m} c_{m}R^{m} +\delta^{\mu}_{\;\nu} \sum_{m} mc_{m} \left(R^{m-1}\right)^{\; ;\sigma}_{;\sigma} -\sum_{m}mc_{m} \left(R^{m-1}\right)^{;\mu}_{\; ;\nu} \nonumber \\
&=:G^{\mu}_{\;\nu}+ H^{\mu}_{\;\nu} = 8\pi T^{\mu}_{\;\nu}\;\; ,\label{eq:eom} \\
&\mbox{with}& \nonumber \\
&G^{\mu}_{\;\nu}:=R^{\mu}_{\;\nu}-\frac{1}{2}R\,\delta^{\mu}_{\;\nu}\;\; . \label{eq:einsterm}
\end{align}
\end{subequations}
In the subsequent text, we consider only $m=-1,\;1$ and $2$. In
such a theory, the vacua ($T^{\mu}_{\;\nu}=0$) are the de Sitter
and anti-de Sitter solutions with effective cosmological
constant, $\Lambda$, given by:
\begin{equation}
\Lambda= \pm \frac{\sqrt{3}}{4}\sqrt{|c_{-1}|} =
\frac{R_{0}}{4}\;\;\; \mbox{ for }\; c_{-1} < 0. \label{eq:vacric}
\end{equation}
Here, $R_{0}$ is the Ricci scalar of the vacuum solution.

\begin{figure}[ht!]
\begin{center}
\includegraphics[bb=0 0 544 435, scale=0.4, clip, keepaspectratio=true]{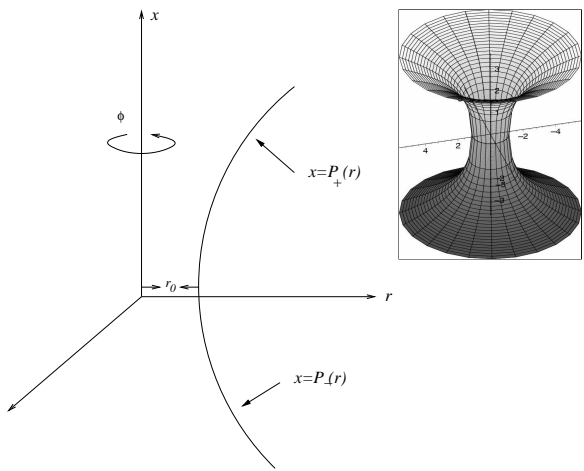}
\caption{{\small Profile curve, $P_{\pm}(r)$, in the $t=$constant
and $\theta=\pi/2$ submanifold. The wormhole is generated via
rotation about the $x$-axis (inset).}} \label{fig:1}
\end{center}
\end{figure}

\qquad In the curvature coordinates often employed in wormhole
studies, the static line element may be written as:
\begin{align}
ds^{2}=& -e^{\gamma(r)}\,dt^{2} + e^{\alpha(r)}\,dr^{2} +r^{2}\,d\theta^{2} +r^{2}\sin^{2}\theta\,d\phi^{2}, \label{eq:curvline} \\
-\infty < &\; t < \infty, \;\;\; 0 < r_{0} \leq r < r_{1}, \;\;\;
0 < \theta < \pi, \;\;\; 0 \leq \phi < 2\pi\; . \nonumber
\end{align}
As we are interested in the region near the throat, we consider
here only the local geometry in this vicinity. The $t=$constant,
$\theta=\pi/2$ profile curve is displayed in figure \ref{fig:1} where
it is embedded in a higher dimensional space, whose extra
coordinate is denoted $x$. Geometric structure is given to the
wormhole manifold by imposing the constraint $x=P_{\pm}(r)$, with
$P_{\pm}(r)$ being the shape function of the profile curve. (The $+$
corresponds to the upper portion of the curve while the $-$
corresponds to the lower portion). The wormhole is obtained by considering
the surface of revolution generated when rotating the profile curve about
the $x$-axis (inset). The lower half of the wormhole is obtained via a
similar construction with a potentially different profile
function. The line element (\ref{eq:curvline}) may be written is
terms of the shape function as:
\begin{equation}
ds^{2}=-e^{\gamma(r)_{\pm}}\,dt^{2} +\left\{1+\left[P_{\pm,r}(r)\right]^{2}\right\}\,dr^{2} +r^{2}\,d\theta^{2} +r^{2}\sin^{2}\theta\,d\phi^{2}\, .
\end{equation}
It is sufficient to study only the upper portion of the wormhole
($P_{+}(r)$) and we therefore suppress the subscript in the
following.

\qquad The restrictions on $P(r)$ are evident from the figure:
\begin{itemize}
\item $P(r)$ must possess positive first derivative, at least in the vicinity of the throat.
\item $P(r)$  must possess negative second derivative near the vicinity of the throat.
\item $P_{,r}(r) \rightarrow +\infty$ as $r \rightarrow r_{0}$ (the throat).
\end{itemize}
It is this last condition which is problematic in the study of
wormhole throats. Although the infinities should cancel out in
calculations of physical quantities involving non-singular
throats, the analysis (both numerical and analytic) is hampered
by the presence of these infinities. A resolution is to switch to
a different coordinate system where such infinities are not
present.

\qquad Another coordinate system, sometimes utilized in wormhole
analysis, is the the proper length gauge where the line element
takes on the form:
\begin{equation}
ds^{2}=-e^{\mu(\rho)}\,dt^{2} +d\rho^{2} +r^{2}(\rho)\,d\theta^{2}
+r^{2}(\rho)\sin^{2}\theta \,d\phi^{2}. \label{eq:propl}
\end{equation}
The coordinate $\rho$ represents the (signed) proper length coordinate in the radial direction:
\begin{equation}
\rho(r)=\pm \int_{r_{0}}^{r}e^{\alpha_{\pm}(r^{\prime})/2}\,dr^{\prime}\; .
\end{equation}
The throat in this chart is located at $\rho=0$. Although
continuous at the throat, the equations of motion are
significantly more complicated in this gauge. There is also the
problem that the profile function is no longer explicit in the
metric which makes the mathematical wormhole engineering a
more difficult task.

\begin{figure}[ht!]
\begin{center}
\includegraphics[bb=0 0 674 409, scale=0.4, clip, keepaspectratio=true]{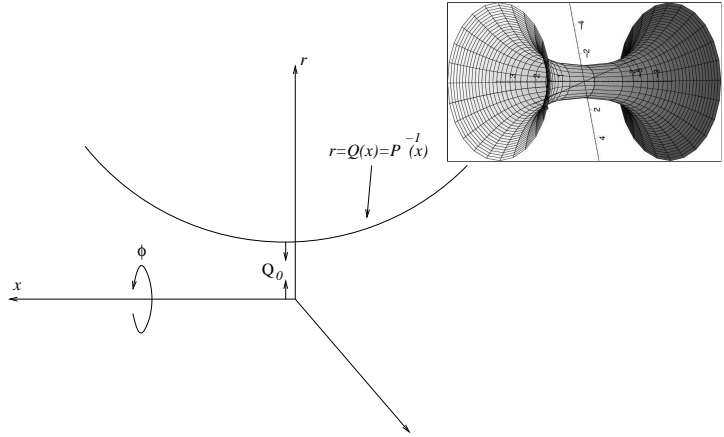}
\caption{{\small Wormhole profile curve in the $t=$constant and
$\theta=\pi/2$ submanifold using the rotated system. The profile
function is given by $r=Q(x)=P^{-1}(x)$ and the radius of the
throat is $Q_{0}$. As before, the wormhole is generated via
rotation about the $x$-axis (inset).}} \label{fig:2}
\end{center}
\end{figure}

\qquad In order to remedy these issues, we instead use a set of
coordinates utilized in \cite{ref:highdworm}. The construction is
briefly summarized here: One begins with the standard curvature
coordinates giving rise to figure \ref{fig:1}. The profile
function, $x=P(r)$ is inverted and instead we consider the new
function given by $r=Q(x)=P^{-1}(x)$. Essentially, we take the
wormhole and rotate it as illustrated in figure \ref{fig:2}. In
this new gauge, the line element takes the form:
\begin{equation}
ds^{2}=-e^{\lambda(x)}\,dt^{2} +
\left\{1+\left[Q_{,x}(x)\right]^{2}\right\} dx^{2}
+Q^2(x)\,d\theta^{2} + Q^{2}(x)\sin^{2}\theta\, d\phi^{2}.
\label{eq:rotlin}
\end{equation}
Notice that this has both the advantage of requiring a single
coordinate chart as well as retaining the profile function in the
metric. The infinite derivative at the throat is replaced with a
vanishing derivative and therefore the metric singularity is
eliminated, causing no trouble for analysis. Further, given the complexity of the equations of motion, we consider here the zero-tidal force class of wormholes as studied in \cite{ref:morthorn}, \cite{ref:9211012}, \cite{ref:tayhis}, \cite{ref:9911099} and others. This amounts to setting $\lambda(x)$ constant, at least near the vicinity of the throat (actually, for the local analysis presented here, $\lambda(x)$ need not be constant so long as the first few derivatives of $\lambda(x)$ vanish at the wormhole throat as not to affect the analytic expansions presented below).

\qquad Here the restrictions for a throat are:
\begin{itemize}
\item The derivative of $Q(x)$ must change sign at the throat. $Q_{,x}(x)<0$ for $x < 0$ and $Q_{,x}(x) > 0$ for $x > 0$.
\item $Q(x)$  must possess positive second derivative near the vicinity of the throat.
\item $Q_{,x}(x)_{|x=0}=0$.
\end{itemize}

\qquad Save for the assumption of static spherical symmetry, the only
other assumption we make regarding the spatial metric is that it is
analytic in the throat region. A general form of profile curve,
suitable for studying any near throat geometry, is given by
\cite{ref:highdworm}:
\begin{subequations}
\begin{align}
Q(x)=&Q_{0}+A^{2}x^{2n}e^{h(x)}, \label{eq:Qprof} \\
Q_{,x}(x)=&A^{2}x^{2n}e^{h(x)}\left[2nx^{-1} +h_{,x}(x)\right], \label{eq:dQprof} \\
Q_{,x,x}=&A^{2}x^{2n}e^{h(x)}\left\{ \left[2nx^{-1} +h_{,x}(x)\right]^{2} +h_{,x,x}(x) -2nx^{-2} \right\}. \label{eq:ddQprof}
\end{align}
\end{subequations}
Here, $A$ and $Q_{0}>0$ are constants, $n$ is a positive integer
sufficiently large to preclude singularities and $h(x)$ is an
arbitrary analytic function of $x$. Higher derivatives are
required for the field equations (\ref{eq:eom}) but no
restriction is placed on them by the requirement that the metric
describe a wormhole throat. The presence of arbitrary functions
allows the modeling of infintely many near-throat geometries. It
is worth noting at this point that the above restrictions may be
relaxed. One could, for example, leave $Q(x)$ as completely
arbitrary, save for the analyticity requirement and the
assumptions on the signs of the first two derivatives. This
yields recondite analysis  but does not change the qualitative
results of this paper.

\section{Weak energy condition}
The weak energy condition is the statement that no
time-like observer measures a local energy density which is negative.
It is usually formulated as:
\begin{equation}
T_{\mu\nu}u^{\mu}u^{\nu} \geq 0 \;\; \forall \;\; \mbox{time-like}\;\; u^{\alpha}\, , \label{eq:WEC}
\end{equation}
the $u^{\alpha}$ being the observer's four-velocity. The solution
for the general WEC in spherical symmetry has been solved in
\cite{ref:ddt}. For the static model presented here, it is
sufficient to study the limiting trajectories in each of the
three principal spatial directions. If the WEC is satisfied for
these trajectories, it will be satisfied for all trajectories.
Enforcing $u^{\alpha}u_{\alpha}=-1$, the limiting cases of
(\ref{eq:WEC}) reduce to the four conditions:
\begin{subequations}
\begin{align}
-T^{0}_{\;0} \geq & 0\;, \\
-T^{0}_{\;0}+T^{i}_{\; i} \geq & 0, \;\;\; \mbox{no summation over }i\;.
\end{align}
\end{subequations}
Using the metric ansatz (\ref{eq:Qprof}) in (\ref{eq:eom}), we expand the
components of the stress-energy tensor about the throat ($x=0$).
The results are summarized here:
\begin{subequations}
\begin{align}
-T^{0}_{\;0} \approx &\frac{1}{4Q_{0}^{4}} \left[4Q_{0}^{2}+\left(1+ 96Q_{0}^{3}A^{2}e^{h(0)}\right) \left(Q_{0}^{6}c_{-1}+8c_{2}\right)\right] \nonumber \\
& +\frac{1}{Q_{0}} 120A^{2}e^{h(0)}(h_{,x})_{x=0} \left(Q_{0}^{6}c_{-1}+8c_{2}\right)\,x +\ldots \;\;\; \mbox{for}\;\;n=2, \label{eq:wec1} \\
-T^{0}_{\;0}\approx& \frac{1}{Q_{0}^{2}} \left(1+\frac{c_{-1}}{4}Q_{0}^{4}
+\frac{2c_{2}}{Q_{0}^{2}}\right) + \frac{(2n)!}{(2n-4)!}A^{2}e^{h(0)}
\left(Q_{0}^{5}c_{-1}+ \frac{8c_{2}}{Q_{0}}\right)\,x^{2n-4} \nonumber \\
& + \ldots \;\;\; \mbox{for} \;\; n>2, \label{eq:wec2}
\end{align}
\end{subequations}
along with:
\begin{subequations}
\begin{align}
-T^{0}_{\;0}+T^{1}_{\;1} \approx& \frac{(2n)!}{(2n-4)!}A^{2}e^{h(0)} \left(Q_{0}^{5}c_{-1} +\frac{8c_{2}}{Q_{0}} \right)\,x^{2n-4} \nonumber \\
&+ \frac{(2n+1)}{(2n-3)} \frac{(2n)!}{(2n-4)!}A^{2}e^{h(0)} (h_{,x})_{|x=0} \left(Q_{0}^{5}c_{-1}+ \frac{8c_{2}}{Q_{0}}\right) \,x^{2n-3} +\ldots , \label{eq:wec3} \\
-T^{0}_{\;0}+T^{2}_{\;2} \approx &\frac{1}{Q_{0}^{2}}-Q_{0}^{2}c_{-1} +4\frac{c_{2}}{Q_{0}^{4}} -\frac{n(2n-1)}{Q_{0}} A^{2}e^{h(0)} \left(2+ \frac{3}{2}Q_{0}^{4}c_{-1} +24 \frac{c_{2}}{Q_{0}^{2}} \right) \, x^{2n-2} \nonumber \\
&+ \ldots \; \approx -T^{0}_{\;0} +T^{3}_{\;3} . \label{eq:wec4}
\end{align}
\end{subequations}
It is easy to see from the above expressions that the WEC, in
principle, may be satisfied in the throat vicinity. The WEC
expression most sensitive to deviations from Einstein gravity is
(\ref{eq:wec3}) (this is the term whose near throat region must
violate the WEC in Einstein gravity).

\qquad The condition for well defined Newtonian limit in
singular non-linear gravity models provided by Dick
\cite{ref:rdick} read that, given $R_{0}>0$:
\begin{equation}
\left| \mathcal{L}(R) \mathcal{L}^{\prime\prime}(R)
\right|_{|R=R_{0}} \ll 1\, . \label{eq:newt2}
\end{equation}
Here primes denote differentiation with respect to $R$ and
$R_{0}$ is the Ricci scalar of the supported vacuum solution. It
was shown that these conditions yield an acceptable Newtonain
limit on length scales $\ll (R_{0})^{-1/2}$. The vacuum of (\ref{eq:eom})
possesses Ricci scalar given by (\ref{eq:vacric}) (for
$m=-1,\;1,\;2$ terms). Therefore, sufficient conditions for (\ref{eq:eom}) to possess
correct Newtonian limit are:
\begin{equation}
c_{-1} < 0, \;\;\; R_{0}=+\sqrt{3 |c_{-1}|}, \;\;\;
c_{2}\approx\frac{1}{3^{3/2}\sqrt{|c_{-1}|}}. \label{eq:dconds}
\end{equation}
The first of these conditions guarantees the existence of a
maximally symmetric gravitational vacuum solution (de Sitter)
whereas the second and third conditions satisfy Dick's restrictions
\cite{ref:rdick}. Therefore, if (\ref{eq:dconds}) holds, WEC violation will not occur as
long as the condition
\begin{equation}
\frac{8}{\sqrt{27}\,|c_{-1}|^{3/2}} > Q_{0}^{6} \label{eq:wecineq}
\end{equation}
is satisfied with $n>1$. For a
theory with only Ricci squared terms the sole restriction is $c_{2}>0$ whereas near throat WEC violation is more severe in the case where only $1/R$ modifications are present.

\begin{figure}[ht!]
\begin{center}
\includegraphics[bb=0 80 493 655, scale=0.4, clip, keepaspectratio=true]{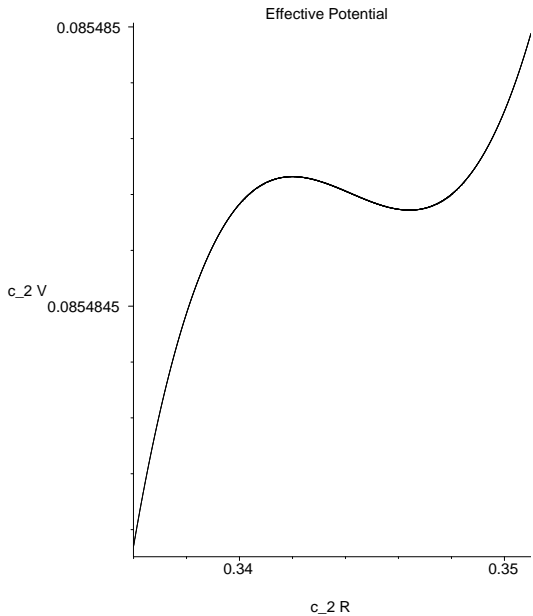}
\caption{{\small Effective potential for stability analysis with
$|c_{-1}|c_{2}^{2}\approx 0.038$.}} \label{fig:pot}
\end{center}
\end{figure}

\qquad Regarding the issue of stability, it is known that the
theory governed by the Lagrangian in (\ref{eq:genaction}) with
$m=-1,\; 1$ and $2$ is equivalent to, in the Einstein frame,
gravitation coupled to a scalar field. The stability of such a
theory has been studied in \cite{ref:nodint} where effective
potential techniques were utilized to analyze the stability. The
effective potential, $V(R)$, is given by \cite{ref:nodint}
\begin{equation}
c_{2}V(R)=
\frac{\left(2|c_{-1}| +c_{2}R^{3}\right)c_{2}R^{3}}{\left(R^{2}+|c_{-1}|+ 2c_{2}R^{3}\right)^{2}}.
\end{equation}
If the last term in (\ref{eq:dconds}) is treated as an equality,
the potential has real stationary points at $R=+1/(3c_{2})$,
$R=0$ and $R=-1/(3c_{2})$ with the product
\begin{equation}
|c_{-1}|c_{2}^{2}=\frac{1}{27}. \label{eq:cprod}
\end{equation}
Under condition (\ref{eq:cprod}) the critical point at $R=+1/(3c_{2})$ is a rising point of
inflection whereas the critical point at $R=-1/(3c_{2})$ is a
minimum. However, the last criterion of (\ref{eq:dconds}) is an
approximate equality and it is easy to see that even a slight
perturbation in the value of (\ref{eq:cprod}) will shift the
inflection point and turn it into a minimum. Thus yielding both weak field
stability and consistent weak field limit. We plot this region
of the potential in figure \ref{fig:pot}. It should be noted that in strong gravitational fields (such as those potentially present near the wormhole throat), $R$ need not be restricted to an interval near this value. However, this analysis is useful in illustrating that the theory possesses acceptable weak field behaviour.

\section{Junction conditions}
We briefly comment here on the patching at the junction away from
the throat (located at $x=x_{*}>0$). As mentioned above, a
patching of the near throat solution to a WEC respecting
space-time could yield a WEC respecting wormhole. There are
several physically acceptable junction conditions one could
employ at the boundary (see, for example, \cite{ref:WECdeb},
\cite{ref:highdworm}, \cite{ref:LLQ}, \cite{ref:Lobo}, \cite{ref:lobo2} for
applications to wormholes in Einstein gravity).

\qquad One condition is that of Synge \cite{ref:synge} which reads:
\begin{equation}
\lim_{\epsilon\rightarrow 0}
\left[T^{\mu}_{\;\nu}\,\hat{n}^{\nu}\right]_{|x=x_{*}-\epsilon} =
\lim_{\epsilon\rightarrow
0}\left[T^{\mu}_{\;\nu}\,\hat{n}^{\nu}\right]_{|x=x_{*}+\epsilon}\;\;.
\label{eq:syngcond}
\end{equation}
Here $\hat{n}^{\nu}$ is the outward pointing unit normal vector of
the junction surface.

\qquad In Einstein gravity, another condition is the Israel-Sen-Lanczos-Darmois (ISLD) junction
conditions \cite{ref:ISLD1}, \cite{ref:ISLD2}, \cite{ref:ISLD3}, \cite{ref:ISLD4} which reads:
\begin{equation}
\lim_{\epsilon\rightarrow
0}\left[K_{\mu\nu}(x_{*}-\epsilon)\right] =
\lim_{\epsilon\rightarrow 0}
\left[K_{\mu\nu}(x_{*}+\epsilon)\right], \label{eq:ISLD}
\end{equation}
with $K_{\mu\nu}$ being the extrinsic curvature tensor of the
junction hyper-surface. It has been shown that, on a time-like
boundary in a spherically symmetric $R$-domain, the above two
conditions are equivalent in Einstein gravity \cite{ref:ddt}.

\qquad The Synge junction condition (\ref{eq:syngcond}) carries
over, in a straight forward manner, into higher derivative
gravity. The ISLD condition (\ref{eq:ISLD}), relying on first
derivatives of the metric, needs to be modified to be valid in
higher derivative theories. One way this could be accomplished is to
consider the boundary terms which arise when applying the
variational principle to (\ref{eq:genaction}). The continuity of
such terms at the junctions will lead to an appropriate junction
condition. Without detailed analysis, it is probably safe to
state that continuity of up to fourth derivatives of the metric
will satisfy such a junction condition (although this restriction
can most likely be relaxed somewhat). This guarantees continuity
of $G^{\mu}_{\;\nu}+H^{\mu}_{\;\nu}$ (or, equivalently here,
$T^{\mu}_{\;\nu}$). The above solution may be patched, for
example, to an anisotropic fluid which is characterized by:
\begin{equation}
T^{\mu}_{\;\nu}=\left( \rho + p_{\perp}\right)u^{\mu}u_{\nu}
+p_{\perp}\delta^{\mu}_{\;\nu}
+\left(p_{\parallel}-p_{\perp}\right)s^{\mu}s_{\nu},
\end{equation}
with $u^{\mu}$ the fluid four-velocity, which is perpendicular to
the vector $s^{\mu}$. The quantities $\rho$, $p_{\perp}$,
$p_{\parallel}$ represent the energy density, transverse pressure
and parallel pressure of the fluid respectively. Other boundaries
further away from the throat, such as a matter-vacuum boundary,
may similarly be considered.

\section{Concluding remarks}
In summary, wormhole throats were studied in a gravitational theory governed by the Einstein-Hilbert Lagrangian supplemented with $1/R$ and $R^{2}$ Ricci scalar terms. The resulting equations of motion was utilized to study energy conditions in the vicinity of the wormhole throat. It was found that the weak energy condition (WEC) may be respected in the throat region in the modified theory. The conditions for a WEC respecting throat are compatible with those required for stability and an acceptable Newtonian limit. Away from the throat, the system cold be joined to WEC respecting matter solutions and, at other junctions, even the vacuum. It would be of interest to study these junction conditions in detail in future work. Other curvature invariant contributions could also be investigated.

\linespread{0.6}
\bibliographystyle{unsrt}

\begin{thebibliography}{10}
{\small

\bibitem{ref:flamm}
L.~Flamm,
\newblock {\em Phys. Z.} {\bf 17} (1916) 448.

\bibitem{ref:einstrose}
A.~Einstein and N.~Rosen,
\newblock {\em Ann. Phys.} {\bf 2} (1935) 242.

\bibitem{ref:wheelfoam}
J.~A.~Wheeler,
\newblock {\em Phys. Rev.} {\bf 48} (1957) 73.

\bibitem{ref:chropro}
S.~W.~Hawking,
\newblock {\em Phys. Rev.} {\bf D46} (1992) 603.

\bibitem{ref:topch}
J.~L.~Friedmann, K.~Schleich and D.~N.~Witt,
\newblock {\em Phys. Rev. Lett.} {\bf 71} (1993) 1486.

\bibitem{ref:horiz}
V.~Frolov and I.~D.~Novikov,
\newblock {\em Phys. Rev.} {\bf D48} (1993) 1607.

\bibitem{ref:sing}
M.~Visser and D.~Hochberg,
\newblock in: {\em Proceedings of the Haifa Workshop on the Internal Structure of Black Holes
and Spacetime Singularities}, (Haifa, Israel, 1997).

\bibitem{ref:morthorn}
M.~S.~Morris and K.~S.~Thorne,
\newblock {\em Am. J. Phys.} {\bf 56} (1988) 395.

\bibitem{ref:WEC1}
M.~S.~Morris, K.~S.~Thorne and U.~Yurtsever,
\newblock {\em Phys. Rev. Lett.} {\bf 61} (1988) 1466.

\bibitem{ref:WECmann}
M.~S.~R.~Delgaty and R.~B.~Mann,
\newblock {\em Int. J. Mod. Phys.} {\bf D4} (1995) 231.

\bibitem{ref:WECkuh}
P.~K.~F.~Kuhfittig,
\newblock {\em Am. J. Phys.} {\bf 67} (1999) 125.

\bibitem{ref:WECdeb}
A.~DeBenedictis and A.~Das,
\newblock {\em Class. Quant. Grav.} {\bf 18} (2001) 1187.

\bibitem{ref:kar}
S.~Kar, N.~Dadhich and M.~Visser,
\newblock {\em gr-qc/0405103} (2004).

\bibitem{ref:khat}
V.~M.~Khatsymovsky,
\newblock {\em Phys.Lett.} {\bf B429} (1998) 254.

\bibitem{ref:eteo}
 E.~Teo,
\newblock {\em Phys.Rev.} {\bf D58} (1998) 024014.

\bibitem{ref:rotberg}
S.~E.~Perez~Bergliaffa, K.~E.~Hibberd,
\newblock {\em gr-qc/0006041} (2000).

\bibitem{ref:kuhax}
P.~K.~F.~Kuhfittig,
\newblock {\em Phys. Rev.} {\bf D67} (2003) 064015.

\bibitem{ref:visbook}
M.~Visser,
\newblock {\em Lorentzian Wormholes: From Einstein to Hawking}. (AIP, New York, 1996).

\bibitem{ref:MTW}
C.~W.~Misner, K.~S.~Thorne and J.~A.~Wheeler,
\newblock {\em Gravitation}. (Freeman, San Francisco, 1973).

\bibitem{ref:nodin1}
S.~Nojiri and S.~D.~Odintsov,
\newblock {\em Mod. Phys. Lett.} {\bf A19} (2004) 627.

\bibitem{ref:allem}
G.~Allemandi, A.~Borowiec and M.~Francaviglia,
\newblock {\em hep-th/0407090} (2004).

\bibitem{ref:starob}
A.~A.~Starobinsky,
\newblock {\em Phys. Lett.} {\bf B91} (1980) 99.

\bibitem{ref:nodint2}
M.~C.~B.~Abdalla, S.~Nojiri and S.~D.~Odintsov,
\newblock {\em Int. Conf. Math. Method. Phys.} (Rio de Janeiro, Brazil, 2004).

\bibitem{ref:cosdyn}
S.~Carloni, P.~K.~S.~Dunsby, S.~Capozziello and A.~Troisi,
\newblock {\em gr-qc/0410046} (2004).

\bibitem{ref:carroll}
S.~M.~Carroll, V.~Duvvuri, M.~Trodden and M.~S.~Turner,
\newblock {\em astro-ph/0306438} (2003).

\bibitem{ref:cct}
S.~Capozziello, S.~Carloni and A.~Troisi,
\newblock {\em Rec. Res. Dev. Astro. Astroph.} RSP/AA/21 (2003).

\bibitem{ref:mengwang}
X.~Meng and P.~Wang,
\newblock {\em Class. Quant. Grav.} {\bf 21} (2004) 951.

\bibitem{ref:vollick}
D.~Vollick,
\newblock {\em Phys. Rev.} {\bf D68} (2003) 063510.

\bibitem{ref:mengwang2}
X.~Meng and P.~Wang,
\newblock {\em gr-qc/0411007}, to appear in {\em Class. Quant. Grav.} (2004).

\bibitem{ref:odin}
S.~Nojiri and S.~D.~Odintsov,
\newblock {\em Phys.Lett.} {\bf B576} (2003) 5.

\bibitem{ref:gzbr}
U.~G\"{u}nther, A.~Zhuk, V.~B.~Bezerra and C.~Romero,
\newblock {\em hep-th/0409112} (2004).

\bibitem{ref:dolgov}
A.~D.~Dolgov and M.~Kawasaki,
\newblock {\em Phys. Lett.} {\bf B573} (2003) 1.

\bibitem{ref:chiba}
T.~Chiba,
\newblock {\em Phys. Lett.} {\bf B575} (2003) 1.

\bibitem{ref:CDDETT}
S.~M.~Carroll, A.~De~Felice, V.~Duvvuri, D.~A.~Easson, M.~Trodden and M.~S.~Turner,
\newblock {\em astro-ph/0410031} (2004).

\bibitem{ref:doming}
A.~E.~Dominguez and D.~E.~Barraco,
\newblock {\em Phys. Rev.} {\bf D70} (2004) 043505.

\bibitem{ref:flan}
E.~E.~Flanagan,
\newblock {\em Class. Quant. Grav.} {\bf 21} (2004) 3817.

\bibitem{ref:rdick}
R.~Dick,
\newblock {\em Gen. Rel. Grav.} {\bf 36} (2004) 217.

\bibitem{ref:GS}
K.~Ghoroku and T.~Soma,
\newblock {\em Phys. Rev.} {\bf D46} (1992) 1507.

\bibitem{ref:hochgeo}
D.~Hochberg and M.~Visser,
\newblock {\em Phys. Rev.} {\bf D56} (1997) 4745.

\bibitem{ref:highdworm}
A.~DeBenedictis and A.~Das,
\newblock {\em Nucl. Phys.} {\bf B653} (2003) 279.

\bibitem{ref:9211012}
T.~A.~Roman,
\newblock {\em Phys. Rev.} {\bf D47} (1993) 1370.

\bibitem{ref:tayhis}
B.~E.~Taylor and W.~A.~Hiscock,
\newblock {\em Phys. Rev.} {\bf D55} (1997) 6116.

\bibitem{ref:9911099}
S.~W.~Kim,
\newblock {\em Grav. Cos.} {\bf 6} (2000) 337.

\bibitem{ref:ddt}
A.~Das, A.~DeBenedictis, N.~Tariq,
\newblock {\em J. Math. Phys.} {\bf 44} (2003) 5637.

\bibitem{ref:nodint}
S.~Nojiri and S.~D.~Odintsov,
\newblock {\em Phys. Rev.} {\bf D68} (2003) 123512.

\bibitem{ref:LLQ}
J.~P.~S.~Lemos, F.~S.~N.~Lobo and S.~Q.~de~Oliveira,
\newblock {\em Phys.Rev.} {\bf D68} (2003) 064004.

\bibitem{ref:Lobo}
F.~S.~N.~Lobo,
\newblock {\em Class. Quant. Grav.} {\bf 21} (2004) 4811.

\bibitem{ref:lobo2}
F.~S.~N.~Lobo,
\newblock {\em gr-qc/0410087} (2004).

\bibitem{ref:synge}
J.~L.~Synge,
\newblock {\em Relativity: The General Theory}
(North-Holland, Amsterdam, 1964).

\bibitem{ref:ISLD1}
W.~Israel,
\newblock {\em Nuovo Cimento} {\bf B44} (1966) 1.

\bibitem{ref:ISLD2}
N.~Sen,
\newblock {\em Ann. Phys.} {\bf 73} (1924) 365.

\bibitem{ref:ISLD3}
C.~Lanczos,
\newblock {\em Ann. Phys.} {\bf 74} 518 (1924) 518.

\bibitem{ref:ISLD4}
G.~Darmois,
\newblock {\em Memorial des Sciences Mathematics XXV}.
(Gauthier-Villars, Paris, 1927).



}

\end{thebibliography}

\end{document}